\documentclass[runningheads]{llncs}
\usepackage{graphicx,subcaption,amsmath}
\usepackage[misc]{ifsym}
\DeclareMathOperator*{\argmax}{arg\,max}
\begin{document}

\mainmatter             

\title{Whole-Sample Mapping of Cancerous and Benign Tissue Properties}

\author{Lydia Neary-Zajiczek\inst{1,2}$^{\textrm{\Letter}}$ \and Clara Essmann\inst{2} \and Neil Clancy\inst{1} \and Aiman Haider\inst{3} \and Elena Miranda\inst{4} \and Michael Shaw\inst{2,5} \and Amir Gander\inst{6} \and Brian Davidson\inst{1} \and Delmiro Fernandez-Reyes\inst{2} \and Vijay Pawar\inst{2} \and Danail Stoyanov\inst{1}}
\index{Neary-Zajiczek, Lydia}
\index{Essmann, Clara}
\index{Clancy, Neil}
\index{Haider, Aiman}
\index{Miranda, Elena}
\index{Shaw, Michael}
\index{Gander, Amir}
\index{Davidson, Brian}
\index{Fernandez-Reyes, Delmiro}
\index{Pawar, Vijay}
\index{Stoyanov, Danail}

\authorrunning{L. Neary-Zajiczek et al.}

\institute{Wellcome/EPSRC Centre for Surgical and Interventional Sciences (WEISS), Charles Bell House, London, UK W1W 7TS\\
\and
UCL TouchLab, Malet Place Engineering Building, London, UK WC1E 7JE\\
\email{Email: lydia.zajiczek.17@ucl.ac.uk}
\and
Department of Cellular Pathology, UCLH, Shropshire House, London, UK \\WC1E 6JA\\
\and
UCL Cancer Research Biobank, Rockefeller Building, London, UK WC1 6JJ\\
\and
National Physical Laboratory, Hampton Road, Teddington, UK TW11 0LW\\
\and
UCL Department of Surgical Biotechnology, Gower Street, London, UK WC1E 6BT}

\maketitle              

\begin{abstract}
Structural and mechanical differences between cancerous and healthy tissue give rise to variations in macroscopic properties such as visual appearance and elastic modulus that show promise as signatures for early cancer detection. Atomic force microscopy (AFM) has been used to measure significant differences in stiffness between cancerous and healthy cells owing to its high force sensitivity and spatial resolution, however due to absorption and scattering of light, it is often challenging to accurately locate where AFM measurements have been made on a bulk tissue sample. In this paper we describe an image registration method that localizes AFM elastic stiffness measurements with high-resolution images of haematoxylin and eosin (H\&E)-stained tissue to within $\pm$1.5 $\mu$m. Color RGB images are segmented into three structure types (lumen, cells and stroma) by a neural network classifier trained on ground-truth pixel data obtained through $k$-means clustering in HSV color space. Using the localized stiffness maps and corresponding structural information, a whole-sample stiffness map is generated with a region matching and interpolation algorithm that associates similar structures with measured stiffness values. We present results showing significant differences in stiffness between healthy and cancerous liver tissue and discuss potential applications of this technique.

\end{abstract}

\section{Introduction}

The development of cancer in otherwise healthy tissue can be detected in numerous ways, however a final diagnosis can only be made after a tissue biopsy or smear, whereby a pathologist inspects stained tissue sections and confirms the presence of cancerous cells. While histopathological analysis remains the ``gold standard" for diagnosis, the macroscopic properties of tumorous tissue are widely known, with surgeons using tactile stiffness information when palpating to locate masses during biopsies or resection surgeries. Attempts have been made to quantify these differences using elastography \cite{Rouviere2017} and tactile sensing \cite{Astrand2017}. Differences have also been measured in the elastic stiffness of single cells cultured from cancerous and healthy tissue using atomic force microscopy (AFM) \cite{Lekka2012,Hayashi2015}, which arise from interactions between cells and the extra-cellular matrix (ECM) \cite{Huang2005,Zajiczek2016}. Given that extra-cellular components contribute significantly to the mechanical properties of the tissue, thicker tissue sections show more significant variation between cancerous and healthy tissue. While AFM measurements are highly sensitive, measuring the stiffness of an entire tissue sample using this method is wholly impractical due to time constraints, and using a small number of measurement sites to characterize bulk tissue properties is problematic as such significant extrapolation is susceptible to sampling errors. Furthermore, accurately localizing where AFM measurements have been made can be difficult as thick unstained tissue sections are highly scattering, resulting in low contrast images when using conventional light microscopy; post-measurement imaging as done in \cite{Plodinec2012} does not allow for any meaningful comparison between measurements and underlying tissue structure.

In this paper we present a combined image registration and propagation method that localizes AFM measurements and hence the estimation of the Young's modulus of thick tissue sections with high-resolution haematoxylin and eosin (H\&E)-stained images of the same section, revealing detailed tissue structure information. Measured tissue properties are propagated through the sample by comparing regions of similar structural content, resulting in whole-sample maps that can be used to draw larger conclusions about relevant nanomechanical differences between cancerous and healthy tissue rather than relying on a small number of measurement sites that are highly localized relative to the size of the sample and not necessarily representative of the tissue as a whole. The tissue property maps generated through this technique allow for detailed analysis of the underlying cellular structure that contributes to measurable changes in macroscopic tissue properties; such information would be of great use in further understanding how cancer develops or in identifying measurable signatures, with implications for early detection of cancerous tissue \cite{Plodinec2012}. 

\section{Methods}

\subsection{Sample Preparation}

Clinical liver tissue samples were obtained, with healthy and cancerous tissue sections acquired from each patient. Malignancy was confirmed through examination of H\&E -stained and fixed thin sections (5 $\mu$m) prior to AFM data acquisition. For each sample undergoing AFM measurements (healthy and cancerous), a thick cryosection (40 $\mu$m) was freshly thawed, adhered to the slide using tissue glue and immersed into physiological buffer for the duration of the measurements, with the buffer solution held in place with a customized 3D printed structure. A custom-built whole slide imaging (WSI) system was used to take a series of low-magnification (4X, 0.16 NA) images of each unstained sample. The low contrast of the thick tissue necessitated the use of a low numerical aperture microscope objective.  The WSI consisted of an Olympus BX63 upright microscope, motorized stages and a large sensor sCMOS RGB camera (PCO edge 3.1c) with a sensor size of 2048 x 1536 pixels, 16-bit (27,000:1) dynamic range and 6.5 $\mu$m square pixel size. Images of the unstained samples were combined using a stitching algorithm employing a Fourier transform-based phase correlation method to find translational offsets followed by small affine transformations and linear blending \cite{stitching}. A typical sized sample (6 mm x 4 mm) required between 5 and 10 low-magnification images to be stitched together.

\subsection{Stiffness Measurement and Registration}

\begin{center}
\begin{figure}
\begin{subfigure}[h]{0.28\textwidth}
\begin{center}
  \includegraphics[width=\linewidth]{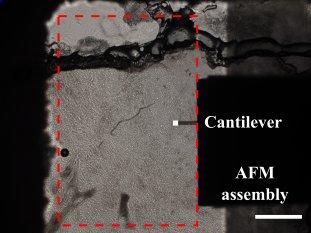}
  \caption{}
  \label{fig:afm_fov}
 \end{center}
\end{subfigure}
\hfill
\begin{subfigure}[h]{0.33\textwidth}
\begin{center}
  \includegraphics[width=\linewidth]{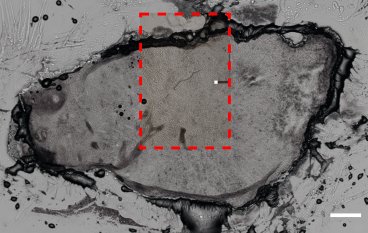}
  \caption{}
  \label{fig:fresh_overlaid}
 \end{center}
\end{subfigure}
\hfill
\begin{subfigure}[h]{0.33\textwidth}
\begin{center}
  \includegraphics[width=\linewidth]{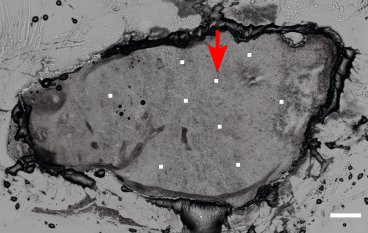}
  \caption{}
  \label{fig:fresh_meas_sites}
 \end{center}
\end{subfigure}
\caption{(a) AFM image with tissue contact point of cantilever indicated by white box (not to scale). Registration FOV indicated by dashed red box. (b)  AFM FOV registered and overlaid on whole sample image. (c) Localized measurement sites (white boxes). Scale bars are all 500 $\mu$m.}
\end{figure}
\end{center}

After determining suitable measurement sites (i.e. areas of the tissue without significant voids or areas that were not firmly affixed to the slide), a JPK Nanowizard 3 Atomic Force Microscope was used in ``force mapping mode" to extract a grid of force-displacement curves over a 10 x 10 $\mu$m area of the sample. Measurements were taken with tipless, soft cantilevers having a 10 $\mu$m borosilicate bead attached (0.03-8 N/m; NSC12 $\mu$Masch). The elastic modulus of the tissue was calculated from the acquired force curves based on the Hertz-Sneddon contact mechanic model \cite{Hertz1882} using JPK analysis software. The same camera and microscope objective from the WSI system was attached to the AFM system for capturing a FOV of the cantilever in place immediately prior to each scan (Fig.~\ref{fig:afm_fov}).
The high dynamic range and large sensor area of the camera allowed for capture of high-contrast detail that greatly improved the robustness of the feature matching and image registration procedures compared to images captured with the camera supplied with the AFM system (sensor size of 1024 x 768 pixels and 8-bit dynamic range). 

To accurately locate the tissue contact point of the cantilever on the AFM FOVs, two images of the cantilever were taken: one at 40X magnification (0.6 NA) with the cantilever bead sharply in focus, where the tissue contact point was visible as the brightest point in the image, and a second at 4X magnification of the cantilever in the same position. These two images were registered using normalized cross correlation. The occlusion of the AFM assembly indicated in Fig.~\ref{fig:afm_fov} resulted in very dark regions which could be thresholded to generate a binary mask showing only the assembly and cantilever, removing all detail of the sample. The cantilever template image with tissue contact point overlaid was also converted to a binary mask, and a feature-matching technique based on speeded up robust features (SURF) was used to estimate any geometric rotation between the cantilever template and AFM FOV masks. Maximum normalized cross correlation was used to register the AFM FOV (Fig.~\ref{fig:afm_fov})
with the unstained image (Fig.~\ref{fig:fresh_overlaid}), localizing the tissue contact point from the cantilever template image of each measurement site to within several microns (Fig.~\ref{fig:fresh_meas_sites}).

The sample was then fixed and stained with H\&E immediately after measurement and imaged with the WSI system at higher magnification (20X, 0.45 NA). Since the samples were significantly thicker than normal tissue sections (40 $\mu$m compared with 5 $\mu$m for regular H\&E stained sections), a volumetric whole-slide scan was carried out, with a 14-plane z-stack spaced by 3 $\mu$m for a total of  42 $\mu$m in focal depth at each field position. A wavelet-based image fusion method was used to generate extended depth of field (EDOF) images containing in-focus structures from all focal planes \cite{Forster2004}. The high-resolution EDOF images were stitched together, the resulting tiled H\&E image was scaled to the same size as the unstained image and the feature-matching normalized cross correlation method described previously was used to register the two whole-sample images. Figs.~\ref{fig:benign_he_meas_sites} and \ref{fig:tumour_he_meas_sites} show localized measurement sites for a healthy and cancerous tissue sample from the same patient overlaid on the corresponding high-resolution tiled H\&E image.

\subsection{Stiffness Propagation}
A combination of methods were used to propagate elastic modulus measurements throughout the sample. The tiled high-resolution H\&E images were first segmented into structural information using a technique based on \cite{Zarella2017}. First, 18 color images (each just over 3 megapixels) from both cancerous and benign tissue samples were manually selected that contained all three relevant structure types: cell nuclei and surrounding cytoplasm, stroma, and lumen or background. These RGB images were converted into HSV color space and all pixel color values (represented as a location in three dimensional space) were grouped into 10 clusters using $k$-means clustering with 10 replicates \cite{Arthur2007}. The distance metric used was squared Euclidean distance, with the $k$ seeds chosen using the $k$-means++ algorithm. Every pixel in each cluster was assigned the cluster centroid as its HSV color value, and the pseudo-color images were displayed for manual structural assignment. Of the 10 pseudo colors, 1 corresponded to lumen, 4 corresponded to cell nuclei and cytoplasm and the remaining 5 were classified as stroma. 

\begin{center}
\begin{figure}[!ht]
\begin{subfigure}[h]{0.47\linewidth}
\begin{center}
  \includegraphics[width=\linewidth]{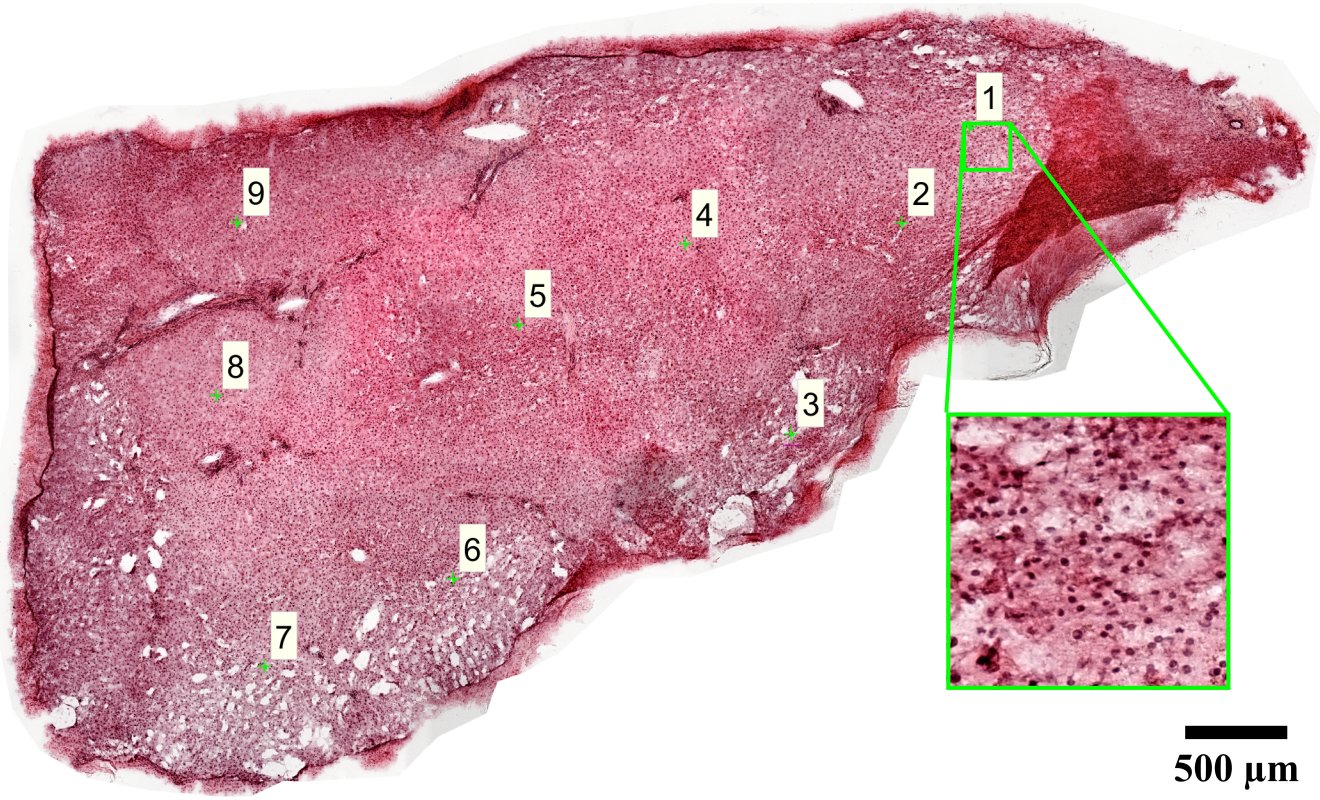}
  \caption{}
  \label{fig:benign_he_meas_sites}
 \end{center}
\end{subfigure}
\hfill
\begin{subfigure}[h]{0.47\linewidth}
\begin{center}
  \includegraphics[width=\linewidth]{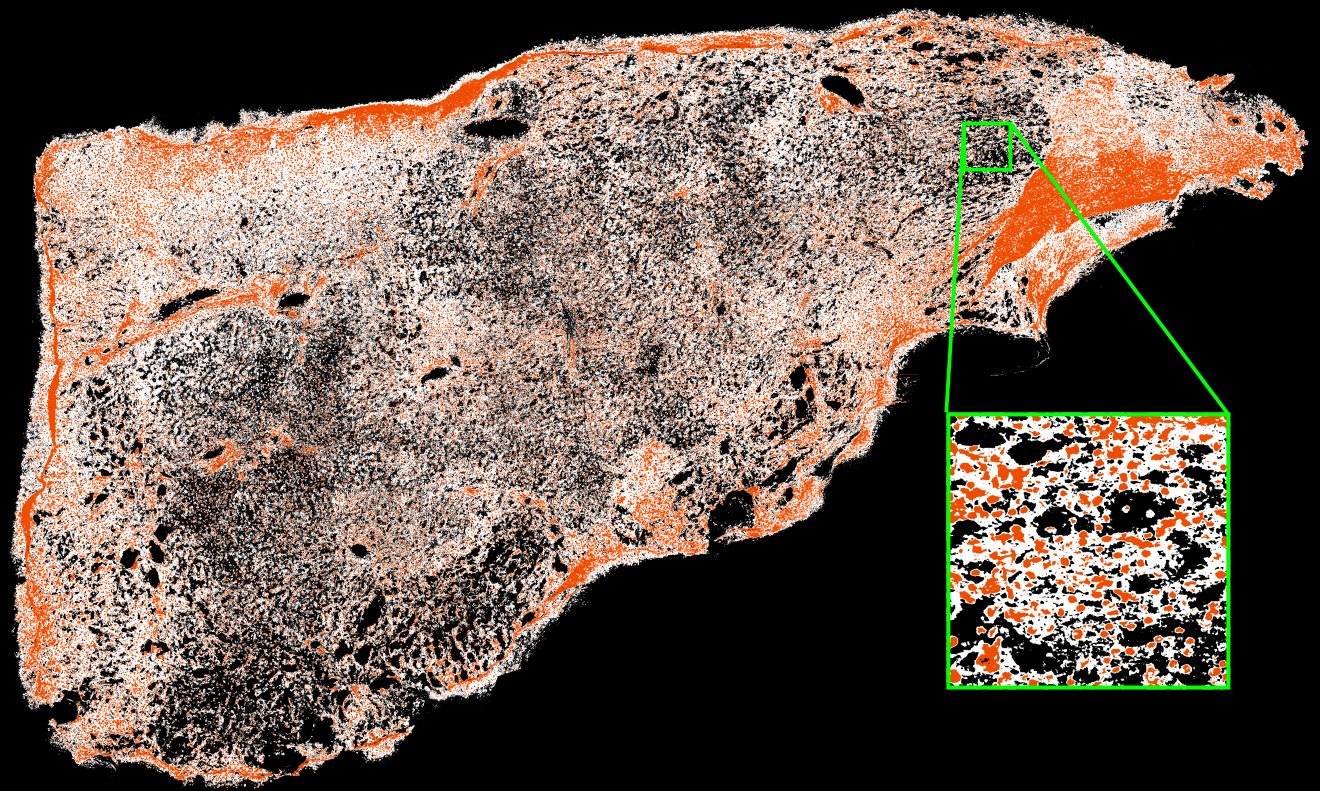}
  \caption{}
  \label{fig:benign_pred}
 \end{center}
\end{subfigure}
\hfill
\begin{subfigure}[h]{0.47\linewidth}
\begin{center}
  \includegraphics[width=\linewidth]{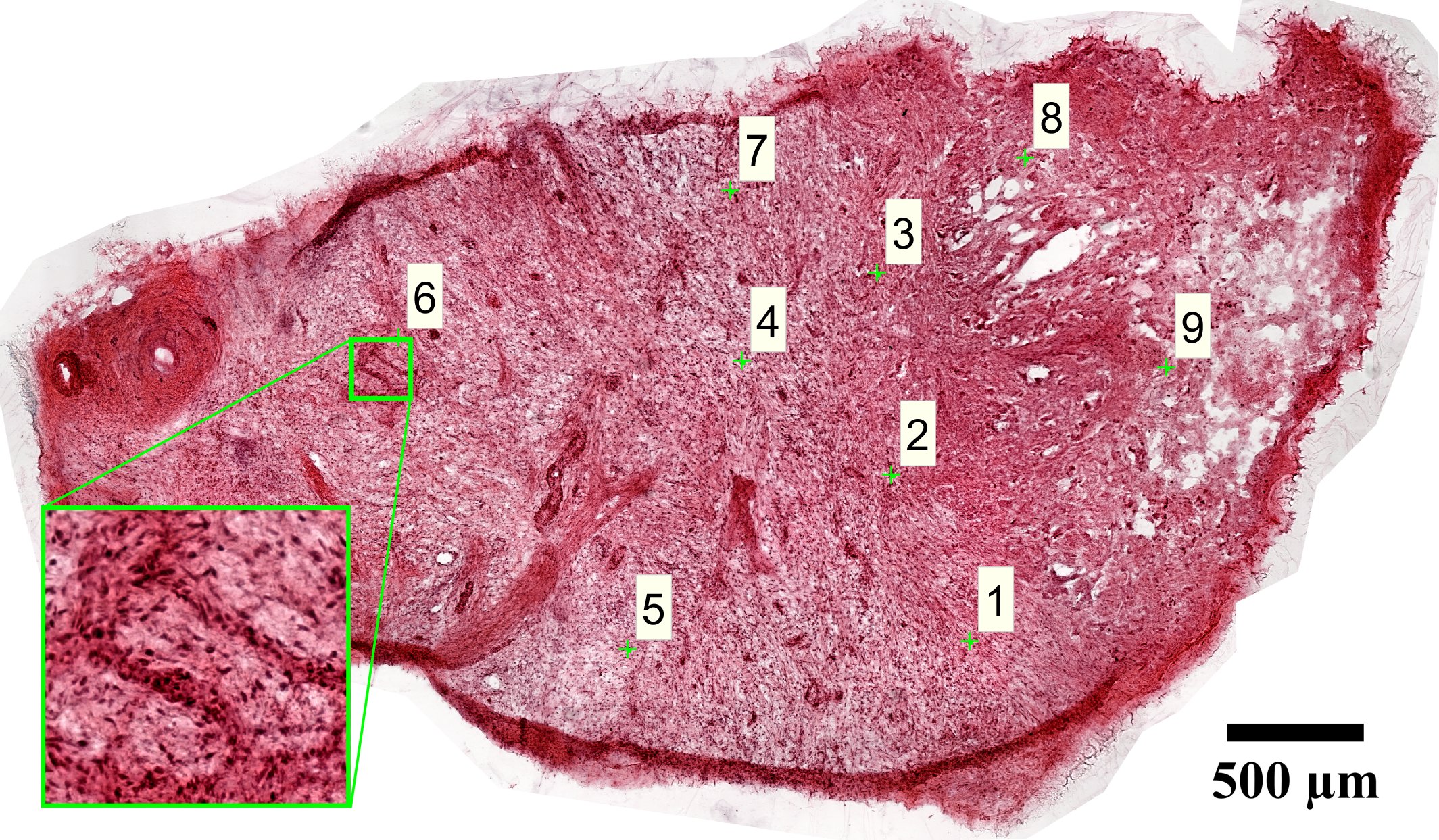}
  \caption{}
  \label{fig:tumour_he_meas_sites}
 \end{center}
\end{subfigure}
\hfill
\begin{subfigure}[h]{0.47\linewidth}
\begin{center}
  \includegraphics[width=\linewidth]{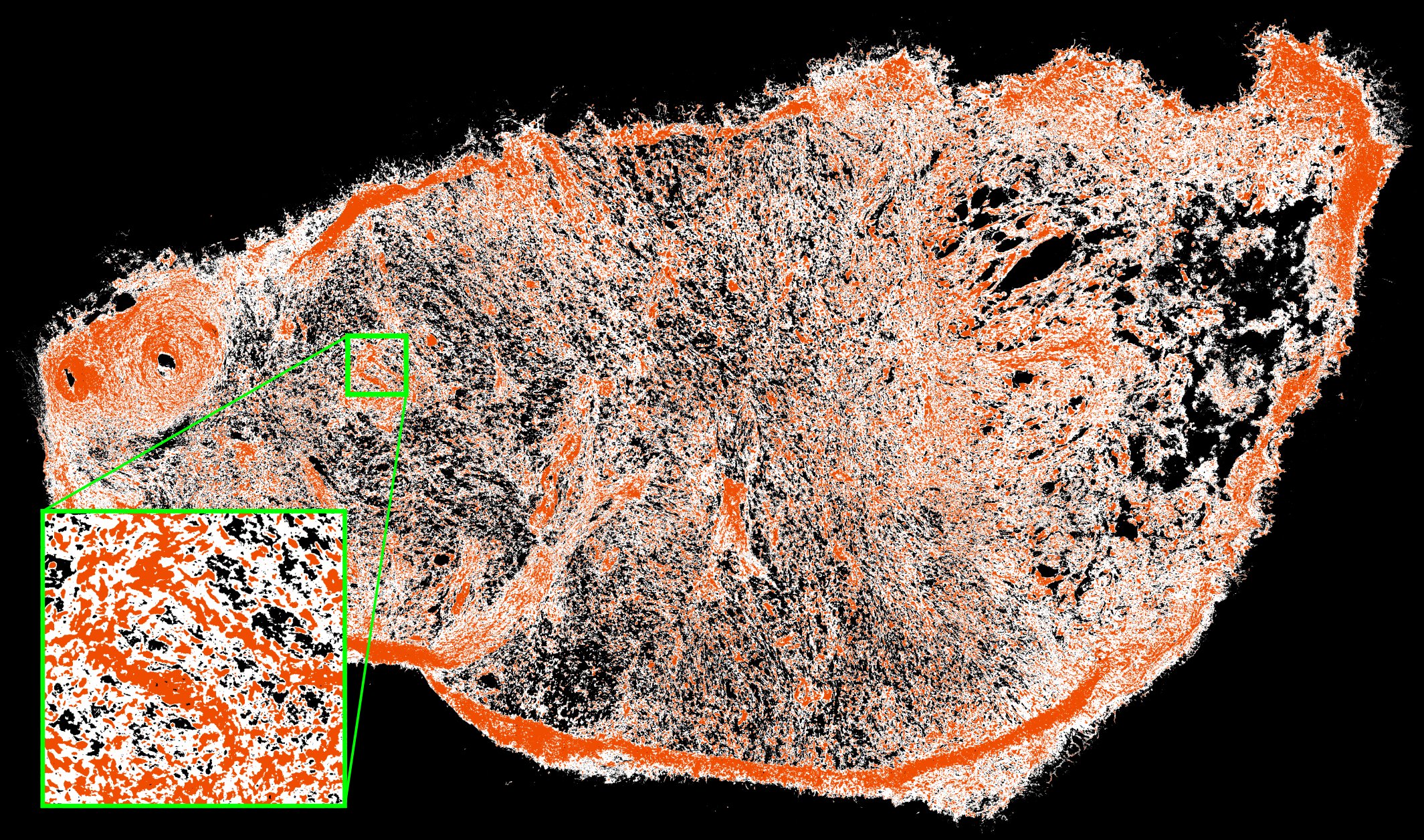}
\caption{}
  \label{fig:tumour_pred}
 \end{center}
\end{subfigure}
\caption{Stitched high-resolution H\&E images of (a) healthy and (c) cancerous samples with measurement sites overlaid. (b) and (d) are corresponding structural information maps of (a) and (b), where white pixels represent stroma, orange represent cells and black represent lumen. Insets are 250 x 250 $\mu$m.}
\label{fig:meas_sites}
\end{figure}
\end{center}

This color assignment generated a training set of approximately 150 million pixels with three input variables (position in 3D HSV space) and three classification labels (lumen, cell or stroma) which was used to train a fully connected neural network classifier in Keras using a Tensorflow backend. The fully connected layer contained 8 hidden nodes and used a rectified linear unit (ReLU) activation function while the output layer used a softmax activation function. The model was compiled using the Adam optimization algorithm with a categorical crossentropy (logarithmic) loss function and accuracy as a performance metric. It was trained with a batch size of 4 times the largest dimension of each image (8192 pixels total) for 20 epochs, achieving maximum training and validation accuracies of 94.5\%. The trained model then predicted the structure type represented by each pixel in the tiled H\&E whole-sample images (Figs.~\ref{fig:benign_pred} and \ref{fig:tumour_pred}). For each measurement location, regions of interest (ROIs) were extracted from both the H\&E and structural images (Figs.~3a and 3b) that were slightly larger than the AFM scan area (Fig.~3c) to account for small errors during registration; ROIs were 13 $\mu$m square in size while the scan areas were 10 $\mu$m square in size. 

\begin{center}
\begin{figure}
\begin{subfigure}[h]{0.3\linewidth}
  \includegraphics[width=\linewidth]{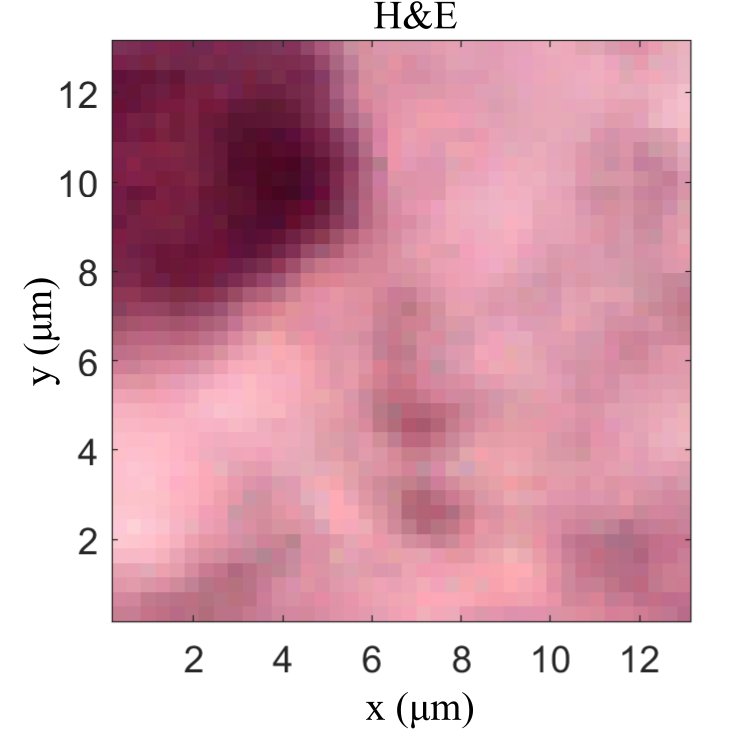}
 \end{subfigure}
\hfill
\begin{subfigure}[h]{0.3\linewidth}
  \includegraphics[width=\linewidth]{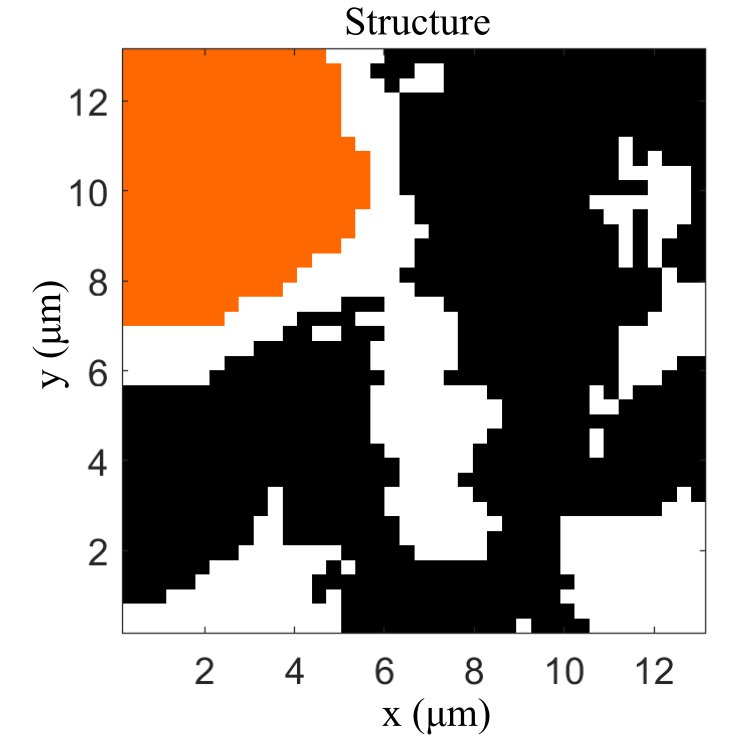}
   \end{subfigure}
\hfill
\begin{subfigure}[h]{0.3\linewidth}
  \includegraphics[width=\linewidth]{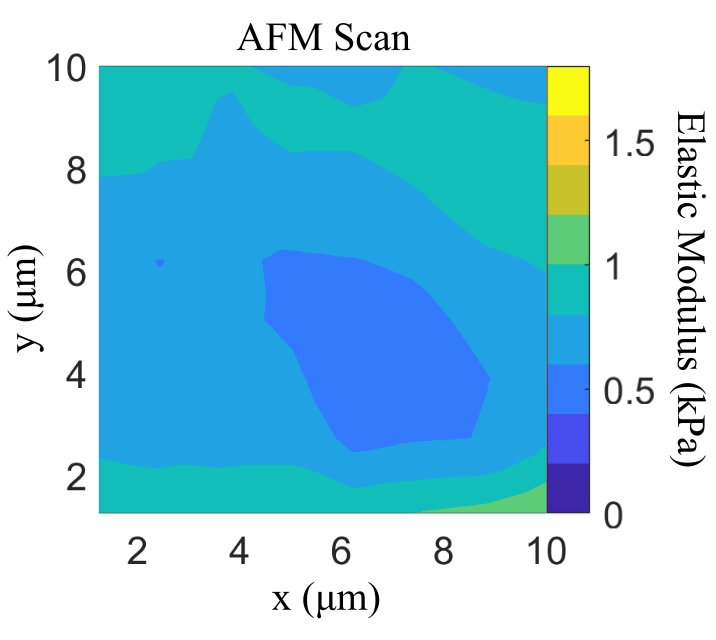}
 \end{subfigure}
 \caption{(a) H\&E image and (b) corresponding structural information map of AFM measurement site \#1 of healthy sample, with (c) showing the measured elastic modulus values of the site. In (b), white pixels represent stroma, orange represent cells and black represent lumen.}
  \label{fig:extracted_fovs}
\end{figure}
\end{center}

The structural information map corresponding to each 10 x 10 $\mu$m AFM measurement site (Fig.~3b) was compared with the structural information map of the entire sample (Fig.~\ref{fig:benign_pred}), using normalized cross correlation (NCC), where  NCC was calculated between the structural information ROIs of all $k$ measurement sites $s_k(x,y)$ and the rest of the structural information map $S(i,j)$, with an arg max selection along $k$ resulting in a maximum correlation map $C(i,j)$:
\[
	C(i,j) = \argmax_k \frac{1}{n}\sum_{x,y} \frac{s_k(x,y) S(i-x,j-y)}{\sigma_{s_k}\sigma_S}
\]
where $i>x$, $j>y$, $n$ is the number of pixels in $(x,y)$ and $\sigma$ denotes standard deviation over $(x,y)$. Note that $C$ is the same size as $S$, i.e. zero padding is not applied at the edges. The correlation map $C$ was then thresholded to identify regions of the sample that are structurally similar to the measurement sites. Each of the thresholded pixels was assigned the average elastic modulus value of the measurement site it is structurally most similar to, resulting in a propagated stiffness map. To better visualize the distribution of stiffness values across the sample, elastic modulus values of unassigned pixels were interpolated using a moving window least-squares approach, but interpolated values were not used in any statistical comparisons.

\section{Results}

\begin{figure}[!ht]
\begin{subfigure}[!ht]{\textwidth}
\centering
  \includegraphics[width=0.84\linewidth]{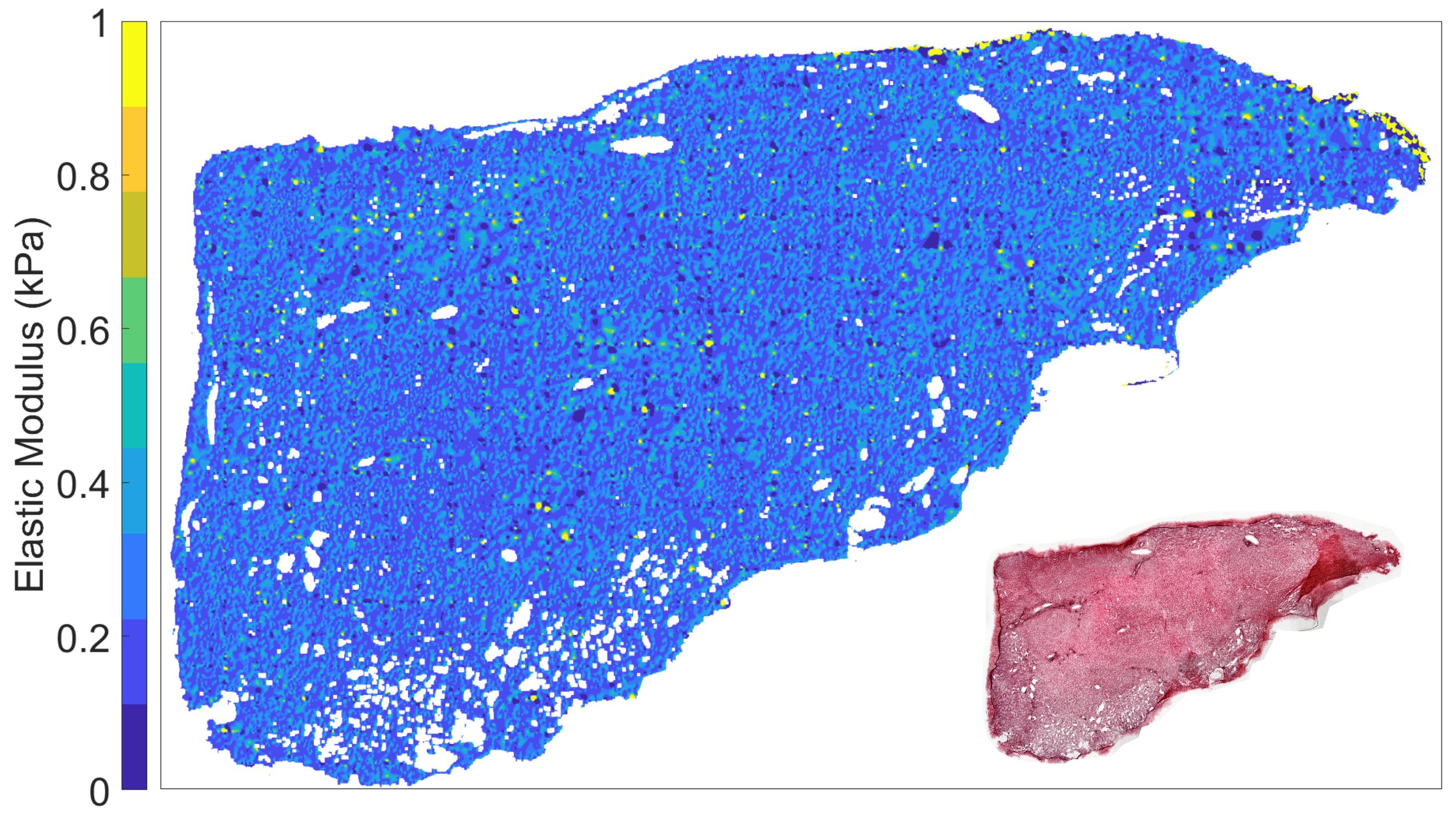}
  \caption{}
  \label{fig:benign_stiff_map}
\end{subfigure}
\begin{subfigure}[!ht]{\textwidth}
\centering
  \includegraphics[width=0.83\linewidth]{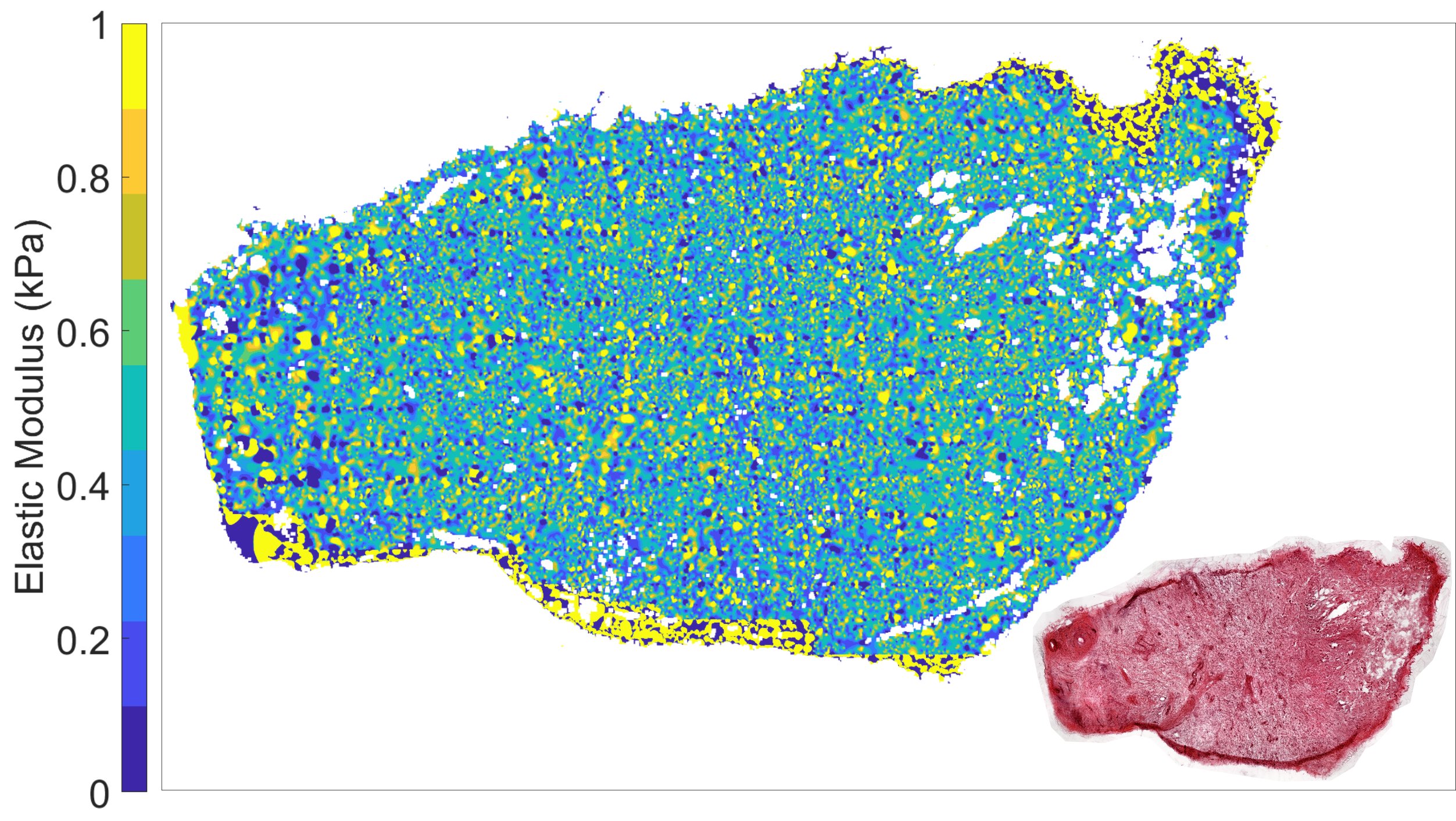}
  \caption{}
  \label{fig:tumour_stiff_map}
\end{subfigure}
\caption{Whole-sample propagated and interpolated maps of (a) healthy and (b) cancerous tissue properties. Maximum values have been truncated to improve visualization. Insets are H\&E stained whole-sample images.}
\label{fig:stiffness_maps}
\end{figure}

The resulting propagated and interpolated tissue property maps are shown in Fig.~\ref{fig:stiffness_maps}. The cancerous sample overall shows a larger average elastic modulus than the healthy sample, with regions of high variation throughout. The average propagated elastic modulus for healthy tissue was 239 $\pm$ 15 Pa, while the cancerous sample's average was 440 $\pm$ 136 Pa. The average of all measured values for each sample was 369 $\pm$ 18 Pa  and 659 $\pm$ 171 Pa for healthy and cancerous tissue, respectively. A t-test performed on 50 randomly selected data points from each measurement set showed a statistically significant difference between propagated elastic modulus values of normal and cancerous tissue ($\bar{p} = 0.004$), however for directly measured values the results of the t-test did not consistently show statistically significant differences. Leave-one-out cross validation was used to test the accuracy of the propagation method, and with the exception of 2-3 well represented measurement sites per sample, removing a single measurement site changed the mean propagated elastic modulus values by less than 5 \%.  

\section{Discussion and Future Work}
The whole-sample tissue property maps presented here are consistent with well-known qualitative observations that cancerous tissue is stiffer and less homogeneous than healthy tissue. The described propagation method is robust, with cross validation results demonstrating that propagation reduces the influence of measured values that may not be representative of the whole sample's structure. Future work will extract quantitative information from the structural maps, such as relative fractions of structural material and nuclear size, and investigate correlations with observed tissue properties. Refinement of the sample mounting technique is also a future consideration, as occasionally AFM measurements can be affected by the sample becoming detached from the slide, reducing the number of valid measurements per sample. Measuring additional samples will result in additional training data for increasing the sensitivity of the structural information classification algorithm, i.e. classifying pixels into more than 3 classes for more detailed and accurate propagation of tissue properties. Finally, the image registration techniques presented here could be applicable in validating potential intraoperative methods for quantitatively measuring the presence of cancer during tissue resectioning for example, using a more robust and scalable stiffness measurement technique than AFM such as direct indentation of tissue using a capacitive force sensor as described in \cite{Zajiczek2016}.

\subsubsection{Acknowledgements.} This work was supported by the Wellcome/EPSRC Centre for Interventional and Surgical Sciences (WEISS) at UCL (203145Z/16/Z), EPSRC (EP/N027078/1, EP/P012841/1, EP/P027938/1, EP/R004080/1) and the European Commission Project-H2020-ICT-24-2015 (Endoo EU Project-G.A. number:688592).

\bibliographystyle{splncs04}
\bibliography{paper2155}

\end{document}